\documentclass[12pt]{article}
\begin{document}
\title{A Note On Fuzzy Space and Topology}
\author{B.G. Sidharth\\
International Institute for Applicable Mathematics \& Information Sciences\\
Hyderabad (India) \& Udine (Italy)\\
B.M. Birla Science Centre, Adarsh Nagar, Hyderabad - 500 063
(India)}
\date{}
\maketitle
\begin{abstract}
We consider the problem of distance between two particles in the
universe, where space is taken to be Liebnizian rather than
Newtonian, this being the present day approach. We then argue that
with latest inputs from physics, it is possible to define such a
distance in a topological sense.
\end{abstract}
\section{Introduction}
We would like to give a statement of the problem in terms of an
actual physical system which is the focus of attention. Increasingly
the Newtonian view of a smooth background space which acts as a
container in which the events in the universe take place is giving
way to the view of Liebnitz in which the contents of the universe
themselves give rise to space \cite{lucas,tduniv}. So we consider
the universe as containing a (finite) number of elementary particles
denoted by $N$. In fact $N$ has been taken to be of the order
$10^{80}.$ The next input is the fact that these $N$ particles are
ill defined to within an extent of the Compton scale
(Cf.refs.\cite{tlsr,tduniv}). All this in present physics leads to a
noncommutative geometry, which is again symptomatic of the non
differentiable nature of spacetime in recent studies (Cf. also
ref.\cite{uof} for a discussion). Indeed space now becomes fuzzy - the
points are ill defined in the minimum intervals \cite{madore1,madore2}.\\
Now, the question is can we define a metric for such a set of
particles or elements?
\section{Topological Considerations}
In earlier work (Cf.ref.\cite{tduniv}) we had argued as follows:
When we talk of a metric or the distance between two "points" or
"particles", a concept that is implicit is that of topological
"nearness" - we require an underpinning of a suitably \index{large
number}large number of "open" sets \cite{simmons}. Let us now
abandon the absolute or background spacetime and consider, for
simplicity, a \index{Universe}Universe (or set) that consists solely
of two particles. The question of the distance between these
particles (quite apart from the question of the observer) becomes
meaningless from the point of view of physical space. Indeed, this
is so for a \index{Universe}Universe consisting of a finite number
of particles. For, we could isolate any two of them, and the
distance between them would have no meaning. We can intuitively
appreciate that we would in fact need distances of intermediate
or more generally, other points.\\
In earlier work\cite{bgsaltaisky,bgsfqv}, motivated by physical
considerations we had considered a series of nested sets or
neighborhoods which were countable and also whose union was a
complete Hausdorff space. The \index{Urysohn Theorem}Urysohn Theorem
was then invoked and it was shown that the space of the subsets was
metrizable.
Let us examine this in more detail.\\
Firstly we observe that in the light of the above remarks, the
concepts of open sets, connectedness and the like reenter in which
case such an isolation of two points would not be possible. More
formally let us define a neighborhood of a particle (or point or
element) $A$ of a set of particles as a subset which contains $A$
and atleast one other distinct element. Now, given two particles (or
points) or sets of points $A$ and $B$, let us consider a
neighborhood containing both of them, $n(A,B)$ say. We require a non
empty set containing at least one of $A$ and $B$ and at least one
other particle $C$, such that $n(A,C) \subset n(A,B)$, and so on.
Strictly, this "nested" sequence should not terminate. For, if it
does, then we end up with a set $n(A,P)$ consisting
of two isolated "particles" or points, and the "distance" $d(A,P)$ is meaningless.\\
We now assume the following property\cite{bgsaltaisky,bgsfqv}: Given
two distinct elements (or even subsets) $A$ and $B$, there is a
neighborhood $N_{A_1}$ such that $A$ belongs to $N_{A_1}$, $B$ does
not belong to $N_{A_1}$ and also given any $N_{A_1}$, there exists a
neighborhood $N_{A_\frac{1}{2}}$ such that $A \subset
N_{A_\frac{1}{2}}
\subset N_{A_1}$, that is there exists an infinite topological closeness.\\
From here, as in the derivation of \index{Urysohn's lemma}Urysohn's
Lemma\cite{munkres}, we could define a mapping $f$ such that $f(A) =
0$ and $f(B) = 1$ and which takes on all intermediate values. We
could now define a metric, $d(A,B) = |f(A) - f(B)|$. We could
easily verify that this satisfies the properties of a metric.\\
With the same motivation we will next deduce a similar result, but
with different conditions. In the sequel, by a subset we will mean a
proper subset, which is also non null, unless specifically mentioned
to be so. We will also consider Borel sets, that is the set itself
(and its subsets) has a countable covering with subsets. We then
follow a pattern similar to that of a Cantor ternary set
\cite{simmons,gullick}. So starting with the set $N$ we consider a
subset $N_1$ which is one of the members of the covering of $N$ and
iterate this process so that
$N_{12}$ denotes a subset belonging to the covering of $N_1$ and so on.\\
We note that each element of $N$ would be contained in one of the
series of subsets of a sub cover. For, if we consider the case where
the element $p$ belongs to some $N_{12\cdots m}$ but not to any
$N_{1,2,3\cdots m+1}$, this would be impossible because the latter
form a cover of the former. In any case as in the derivation of the
\index{Cantor set}Cantor set, we can put the above countable series
of sub sets of sub covers in a one to one correspondence with
suitable sub intervals of a real interval
$(a,b)$.\\
{\large \bf{Case I}}\\
If $N_{1,2,3\cdots m} \to$ an element of the set $N$ as $m \to
\infty$, that is if the set is closed, we would be establishing a
one to one relationship with points on the interval $(a,b)$ and
hence could use the
metric of this latter interval, as seen earlier.\\
{\large \bf{Case II}}\\
It is interesting to consider the case where in the above iterative
countable process, the limit does not tend to an element of the set
$N$, that is set $N$ is not closed and has what we may call singular
points. We could still truncate the process at $N_{1,2,3\cdots m}$
for some $m > L$ arbitrary and establish a one to one relationship
between such truncated subsets and arbitrarily small intervals in
$a,b$. We could still speak of a metric or distance between two
such arbitrarily small intervals.\\
This case which may be termed "Fuzzy Topology", is of interest
because of our description of \index{elementary particles}elementary
particles in terms of fuzzy spacetime (Cf. also ref.\cite{cu}),
where we have a length of the order of the \index{Compton
wavelength}Compton wavelength as seen in the previous section,
within which \index{spacetime}spacetime as we know it breaks down.
Such cut offs lead to a non commutative geometry and what may be
called fuzzy
spaces\cite{tduniv,uof,madore1,madore2}.\\
To put it another way, in view of the fact that the number of
particles or points or elements of the universe, which is the set
under consideration is finite, it may be pointed out that the
sequence $N_{1,2,3\cdots m}$ would terminate for some finite $L_1$,
let us say. We could still identify these sub sets, sub covers etc.
with open sets, sub sets etc. of some manifold $M$ which has
interminable sequences, so that a metric by the previous arguments
would exist for $M$. We could then truncate the sub sets,
$N_{1,2,3\cdots m}$ for some $L_1$ as above, keeping in view the
fact that each particle is not defined to within its Compton scale,
and the fact that there are a finite number of particles. There
would be thus a sequence $L_1,L_2,\cdots L_N$ for each of the $N$
particles of the universe. We could consider $L = inf (L_1,L_2$
etc), $L$ representing some fundamental irreduceable scale like the
Compton scale. What happens within the scale is in any case ill
defined characteristic of the
fuzzyness of space.\\
We could reformulate the above problem in the following simpler
fashion. Let us take a set $S$ of $N$ open sets and put the $N$
particles of the universe in a one to one correspondence with the
open sets of $S$. As $N$ is finite, there are a finite number of
geometric transformations in the spirit of Smale's original
transformations, which take $S$ to $S'$ such that $S'$ is a
metrizeable set of open sets. In this process any overlaps of the
points of the images of open sets of $S$, is not important in view
of the fuzzyness associated with the Compton scale of the $N$
particles. So we have a one to one correspondence between the $N$
particles of the universe with the sub sets of a metrizeable set
$S'$, which in fact is all that is required.\\
To give a simple example, let the universe consist of just two
particles, $A_1$ and $A_2$. By earlier arguments the distance
between them would be ill defined (in a physical sense, not an
artificial mathematical sense). We associate with $A_1$ and $A_2$
two open sets $s_1$ and $s_2$ of $S$. Then we consider a
transformation $S$ to $S'$ such that the images $s'_1$ and $s'_2$
form a set that satisfies standard properties such as connected so
that we can define a distance or metric. This is taken over to be a
"distance" between $A_1$ and $A_2$ remembering that in any case the
points within a fundamental interval around them are
indeterminate.\\
Finally, it may be remarked that if $N$ is countable (infinite) then
there are elementary topologies that can be formed
(Cf.ref.\cite{lips}).
\section{Remarks}
We would like to reiterate the following. Interestingly, we usually
consider two types of infinite sets - those with cardinal number $n$
corresponding to countable \index{infinities}infinities, and those
with cardinal number $c$ corresponding to a continuum, there being
nothing in between \cite{gullick}. This is the well
known but unproven \index{Continuum hypothesis}Continuum hypothesis.\\
What we have shown with the above process is that it is possible to
conceive of an intermediate possibility
with a cardinal number $n^p, p > 1$.\\
In the above considerations three properties are important: Firstly
the set must be closed i.e. it must contain all its limit points.
Secondly, it must be perfect i.e. in addition, each of its points
must be a limit point. Finally it must be disconnected i.e. it
contains no non null open intervals. Only
the first was invoked in Case I.\\
We notice that there is the holistic feature. A metric emerges by
considering large encompassing sets. Finally, we could deviate from
a strict mathematical analysis and introduce an element of physics.
We could say that a point or particle $B$ would be in a neighborhood
of another point or particle $A$, only if $A$ and $B$ interact".
Thus the universe would consist of a network of "interacting"
particles, reminiscent of the Feynman-Wheeler perfect absorber model
encountered.

\end{document}